\documentclass{jpsj-suppl}
\usepackage{txfonts} 
\usepackage{graphicx}

\title{Extending the Eikonal Approximation to Low Energy}

\author{Pierre \textsc{Capel}$^{1}$, Tokuro \textsc{Fukui}$^{2}$ and Kazuyuki \textsc{Ogata}$^{2}$}

\inst{$^{1}$Physique Nucl´eaire et Physique Quantique C.P. 229,
Universit\'e Libre de Bruxelles (ULB), B-1050 Brussels, Belgium\\
$^{2}$Research Center for Nuclear Physics, Osaka University, Ibaraki, Osaka 567-0047, Japan}

\email{pierre.capel@ulb.ac.be}

\recdate{\today}

\abst{E-CDCC and DEA, two eikonal-based reaction models are compared to CDCC at low energy (e.g.\ $20A$MeV) to study their behaviour in the regime at which the eikonal approximation is supposed to fail.
We confirm that these models lack the Coulomb deflection of the projectile by the target.
We show that a hybrid model,  built on the CDCC framework at low angular momenta and the eikonal approximation at larger angular momenta gives a perfect agreement with CDCC.
An empirical shift in impact parameter can also be used reliably to simulate this missing Coulomb deflection.}

\kword{eikonal approximation, nuclear-reaction theory, Coulomb deflection, halo nuclei}

\begin{document}
  \newcommand {\nc} {\newcommand}
  \nc {\beq} {\begin{eqnarray}}
  \nc {\eeq} {\nonumber \end{eqnarray}}
  \nc {\eeqn}[1] {\label {#1} \end{eqnarray}}
  \nc {\Sec} [1] {Sec.~\ref{#1}}
  \nc {\tbl} [1] {Table~\ref{#1}}
  \nc {\fig} [1] {Fig.~\ref{#1}}
  \nc {\Eq} [1] {Eq.(\ref{#1})}
  \nc {\Ref} [1] {Ref.~\cite{#1}}
  \nc {\eq} [1] {(\ref{#1})}
  \nc {\ex} [1] {$^{#1}$}
  \nc {\ve} [1] {\mbox{\boldmath $#1$}}
  \nc {\flim} [2] {\mathop{\longrightarrow}\limits_{{#1}\rightarrow{#2}}}

\maketitle
\section{Introduction}
The development of radioactive-ion beams in the mid-80s has enabled us to explore the nuclear landscape far from stability.
This led to the discovery of unexpected nuclear features like halos.
Halo nuclei are neutron-rich nuclei that exhibit a much larger matter radius than their isobars \cite{Tan96}.
This exotic property is qualitatively understood as resulting from the low binding energy for one or two neutrons of these nuclei.
Thanks to this loose binding, the valence neutrons have a significant probability of presence at a large distance from the other nucleons.
They thus form a sort of halo around the core of the nucleus, which explains the unexpected large size of the nucleus.
Being located close to the dripline, these nuclei exhibit a very short half life and hence cannot be studied through usual spectroscopic technique.
One must then rely on indirect techniques, such as reactions to infer structure information about these nuclei.

Breakup reaction is one of the mostly used tools to study halo nuclei \cite{Fuk04,Nak09}.
In this reaction, the core-halo structure is dissociated through the interaction with a target.
To extract valuable information from breakup observables, accurate reaction models coupled to realistic descriptions of the projectiles are needed.
Various models have been developed to analyse these reactions (see \Ref{BC12} for a recent review)
among which the Continuum Discretised Coupled Channel model (CDCC)\cite{Kam86,TNT01} and eikonal-based models such as the eikonal-CDCC (E-CDCC) \cite{Oga03, Oga06} and the Dynamical Eikonal Approximation (DEA) \cite{BCG05,GBC06}.
In a recent publication, CDCC and DEA have been compared for the breakup of \ex{15}C, a one-neutron halo nucleus, on Pb at intermediate ($70A$MeV) and low ($20A$MeV) energy \cite{CEN12}.
At intermediate energy, both models compare very well with each other and they are in excellent agreement with experiment.
At lower energy, the eikonal approximation fails leading to angular distributions for breakup shifted at too forward an angle.
The analysis in \Ref{CEN12} shows that it is due to the lack of Coulomb deflection in the eikonal approximation.
In the present contribution, we analyse corrections to the eikonal-based models that efficiently take account for this Coulomb deflection.

\section{Theoretical framework}
The models we consider in this work assume a two-body description of the projectile: an inert core $c$ to which a halo nucleon $f$ is loosely bound.
The  Hamiltonian modelling the internal structure of such two-body nuclei reads
\beq
H_0=T_r+V_{cf}(\ve{r}),
\eeqn{e1}
where $\ve{r}$ is the $c$-$f$ relative coordinate.
The $c$-$f$ relative motion is described by the eigenstates of $H_0$
\beq
H_0 \Phi_{lm}(\epsilon,\ve r)=\epsilon\ \Phi_{lm}(\epsilon,\ve r),
\eeqn{e2}
where $l$ is the $c$-$f$ relative angular momentum and $m$ is its projection
(in this study we neglect the spins of the fragments for simplicity).
The negative-energy eigenstates correspond to the bound states of the system.
In the following the ground state of energy $\epsilon_0$ is denoted by $\Phi_0$.
The positive-energy states describe the $c$-$f$ continuum, i.e.\ the broken up projectile.

The target $T$ is usually described as a structureless particle and its interaction with the projectile constituents $c$ and $f$ is simulated by the optical potentials $V_{cT}$ and $V_{fT}$, respectively.
Within this framework, studying the $P$-$T$ collision reduces to solving the three-body Schr\"odinger equation
\beq
\left[T_R+H_0+V_{cT}(\ve{R}_{cT})+V_{fT}(\ve{R}_{fT})\right]\Psi(\ve{r},\ve{R})=E_T\Psi(\ve{r},\ve{R}),
\eeqn{e3}
where $\ve R$ is the coordinate of the projectile centre of mass relative to the target, 
while $\ve{R}_{cT}$ and $\ve{R}_{fT}$ are the $c$-$T$ and $f$-$T$ relative coordinates, respectively.
\Eq{e3} is solved with the condition that the projectile, initially in its ground state, is impinging on the target:
\beq
\Psi(\ve{r},\ve{R})\flim{Z}{-\infty}e^{iKZ+\cdots}\Phi_0(\ve{r}),
\eeqn{e5}
where $K$ is the initial $P$-$T$ wave number, which is related to the total energy $E_T=\hbar^2K^2/2\mu_{PT}+\epsilon_0$, with $\mu_{PT}$ the projectile-target reduced mass.

\subsection{CDCC}
In the Continuum Discretised Coupled Channel method (CDCC), \Eq{e3} is solved expanding the three-body wave function $\Psi$ upon the projectile eigenstates $\Phi_{lm}(\epsilon)$ \cite{Kam86,TNT01}
\beq
\Psi(\ve{r},\ve{R})=\sum_i\chi_i(\ve{R})\Phi_i(\ve{r}),
\eeqn{e6}
where $i$ stands for $l$, $m$ and $\epsilon$.
This leads to a set of coupled equations
\beq
\left[T_R+\epsilon+V_{ii}\right]\chi_i+\sum_{j\ne i}V_{ij}\chi_j=E_T\chi_i,
\eeqn{e7}
where $V_{ij}=\langle\Phi_i|V_{cT}+V_{fT}|\Phi_j\rangle$ are coupling the various channels.

Since the model aims at describing the breakup of the projectile, the expansion \eq{e6} must include a reliable description of the continuum of the projectile.
To be tractable, this description is obtained by discretisation of the continuum \cite{Yah86}.
To reach convergence a large number of states $\Phi_i$ must be included in the expansion \eq{e6}, especially at large beam energies.
This requires large computer power, and limits the extension of CDCC to descriptions of the projectile beyond the simple two-body model presented here.
It is therefore interesting to develop reliable approximations that are less computationally expensive.

\subsection{Eikonal framework}
The eikonal approximation \cite{Glauber} assumes that at sufficiently high beam energy the $P$-$T$ relative motion does not deviate much from the incoming plane wave \eq{e5}.
To simplify the resolution of \Eq{e3}, one can factorise that plane wave out of the three-body wave function
\beq
\Psi(\ve r,\ve R)=e^{iKZ}\widehat\Psi(\ve r,\ve R)
\eeqn{e9}
to obtain a function $\widehat \Psi$ that varies smoothly with $\ve R$.
Neglecting the second-order derivatives of $\widehat\Psi$ in comparison to its first-order derivatives leads to the following equation
\beq
i\hbar v \frac{\partial}{\partial Z}\widehat{\Psi}(\ve{r},\ve{b},Z)=[H_0-\epsilon_0+V_{cT}+V_{fT}]
\widehat{\Psi}(\ve{r},\ve{b},Z),
\eeqn{e10}
which has to be solved with the boundary condition $\widehat\Psi(\ve{r},\ve{R})\flim{Z}{-\infty}\Phi_0(\ve r)$.
This equation is simpler to solve than the set of coupled equations \eq{e7} and hence requires less computational efforts to be solved.
The Dynamical Eikonal Approximation (DEA) solves this equation expanding $\widehat\Psi$ over a three-dimensional spherical mesh \cite{BCG05,GBC06}.

\section{Coulomb deflection}
\begin{figure}
\center
\includegraphics[trim=1.5cm 18cm 7cm 1.7cm, clip=true, width=8.3cm]{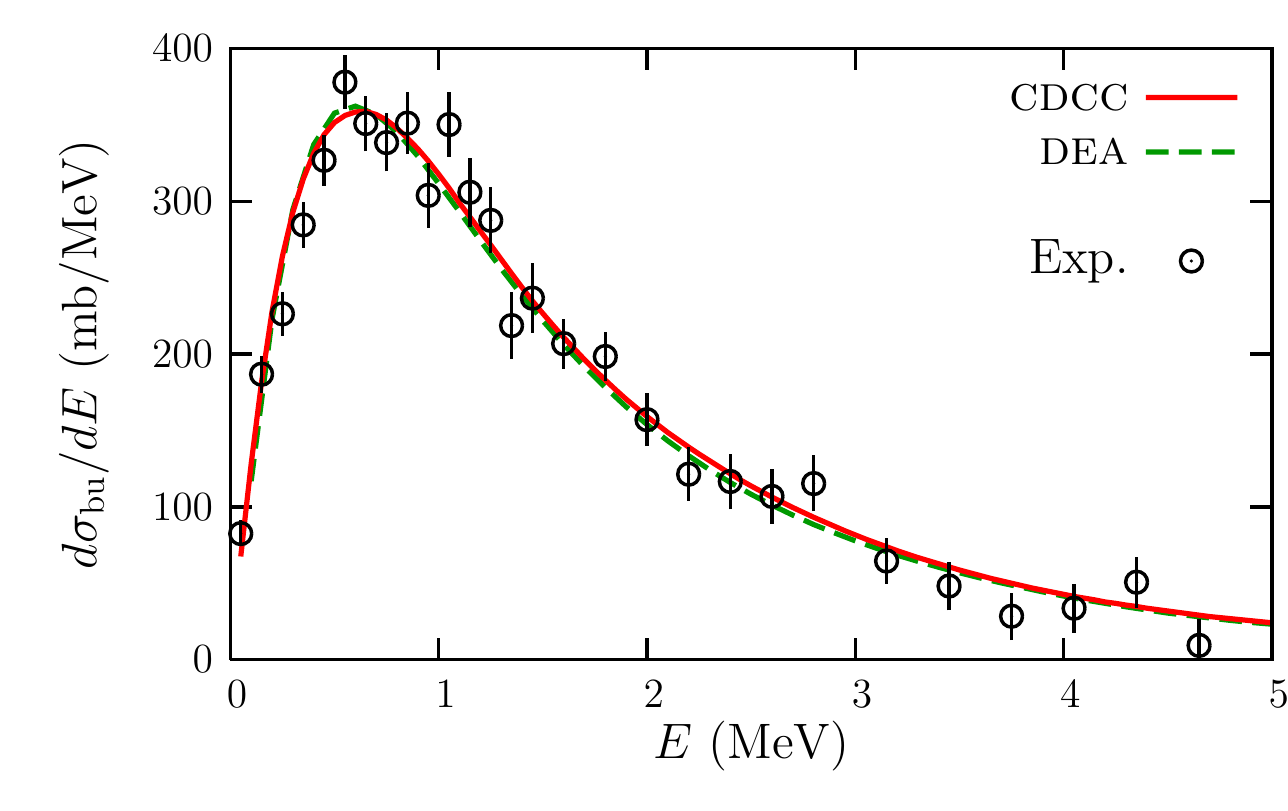}
\includegraphics[trim=2.5cm 18cm 7cm 1.7cm, clip=true, width=7.7cm]{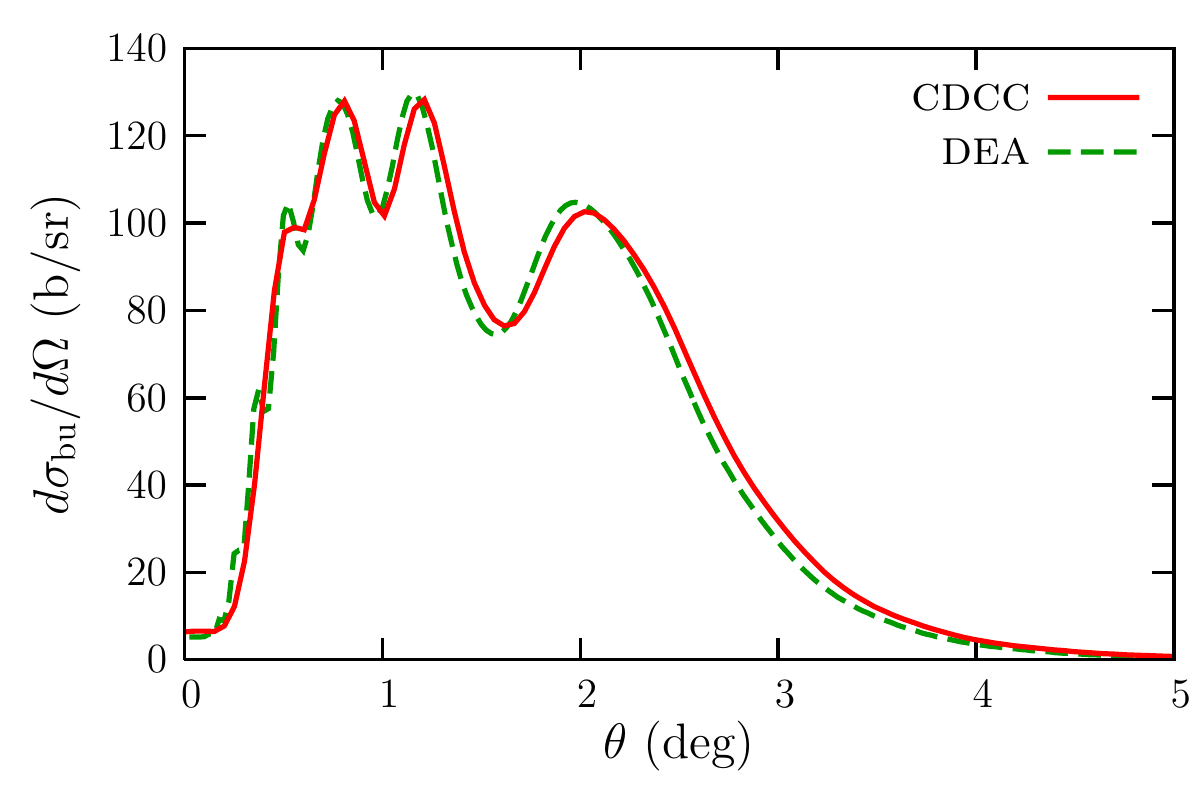}
\caption{Comparison of CDCC and DEA for \ex{15}C impinging on Pb at $68A$MeV \cite{CEN12}.
Energy distribution (left) and angular distribution (right), data from Ref.~\cite{Nak09}.}\label{f1}
\end{figure}

In \Ref{CEN12}, CDCC and DEA have been confronted to each other for the Coulomb breakup of \ex{15}C on Pb at $68A$MeV and $20A$MeV.
At $68A$MeV, both models give very similar cross sections (see \fig{f1}), they also agree very well with the experimental data \cite{Nak09}.

At $20A$MeV, the eikonal approximation no longer holds, and we observe a significant shift of the angular cross section towards forward angles of the DEA compared to CDCC (see \fig{f2} left).
This is understood as a lack of Coulomb deflection within the eikonal framework \cite{CEN12}.
Within a semiclassical viewpoint, \Eq{e10} describes a projectile moving along a straight-line trajectory, hence ignoring the deflection experienced by a realistic projectile subject to the Coulomb field of the target.

To account for this Coulomb deflection, we compare the DEA and full CDCC calculations to the predictions of the eikonal-CDCC (E-CDCC) \cite{Oga03, Oga06}, a model that is based on the eikonal approximation, but that includes part of the projectile-target Coulomb interaction in its expansion (see \Eq{e11}), and that can be easily extended into a hybrid solution, which gives similar results as the full CDCC, although it is much less computationally expensive \cite{Oga03, Oga06}.

\subsection{Eikonal-CDCC}
The Eikonal-CDCC (E-CDCC) solves \Eq{e3} expanding $\Psi$ upon $H_0$ eigenstates $\Phi_{i} ({\ve r})$
considering, as in CDCC, a discretised continuum
\beq
\Psi({\ve r},{\ve R})
=\sum_{i}\xi_{i}({\ve b},Z)  \Phi_{i} ({\ve r})e^{i\left\{K_i Z+\eta_{i}\ln\left[K_{i}R-K_{i}Z\right]\right\}}
\eeqn{e11}
Following the eikonal approximation, this leads to a set of coupled equations, yet simpler to solve than the full CDCC equations \eq{e7}:
\beq
\frac{\partial}{\partial Z}\xi_{i}({\ve b},Z)=
\frac{1}{i\hbar v_i(R)}\sum_{i'}
\mathcal{F}_{ii'}({\ve b},Z)
\xi_{i'}({\ve b},Z)
e^{i(K_{i'}-K_i)Z}
\mathcal{R}_{ii^{\prime}}(b,Z),
\eeqn{e12}
with the coupling potential
\beq
\mathcal{F}_{ii'}({\ve b},Z)
=
\left\langle \Phi_{i}
\left\vert
V_{cT}+V_{fT}-V_{\rm C}
\right\vert
\Phi_{i'} \right\rangle_{\ve r}.
\eeqn{e13}
In this way, E-CDCC takes proper account of the energy conservation since the projectile-target relative velocity $v_i$ varies with the $P$-$T$ distance $R$ and the projectile state $\Phi_i$.
The factor $\mathcal{R}_{ii^{\prime}}(b,Z)
=
\frac{\left(K_{i'}R-K_{i'}Z\right)  ^{i\eta_{i'}}}
{\left(K_{i}R-K_{i}Z\right)  ^{i\eta_{i}}}$
accounts for part of the Coulomb distortion that is missing in the DEA.

\begin{figure}[t]
\includegraphics[trim=2.5cm 17.2cm 7cm 2cm, clip=true, width=8cm]{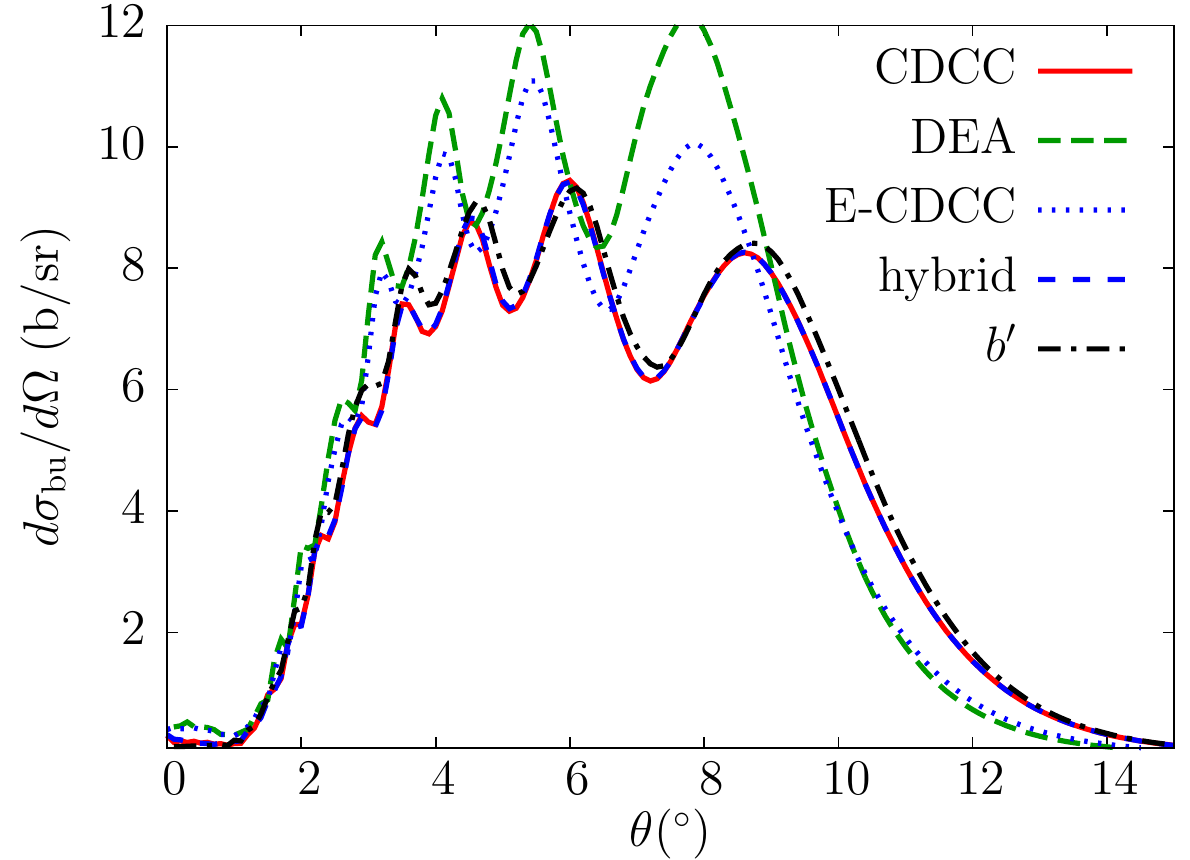}
\includegraphics[trim=2.5cm 17.2cm 7cm 2cm, clip=true, width=8cm]{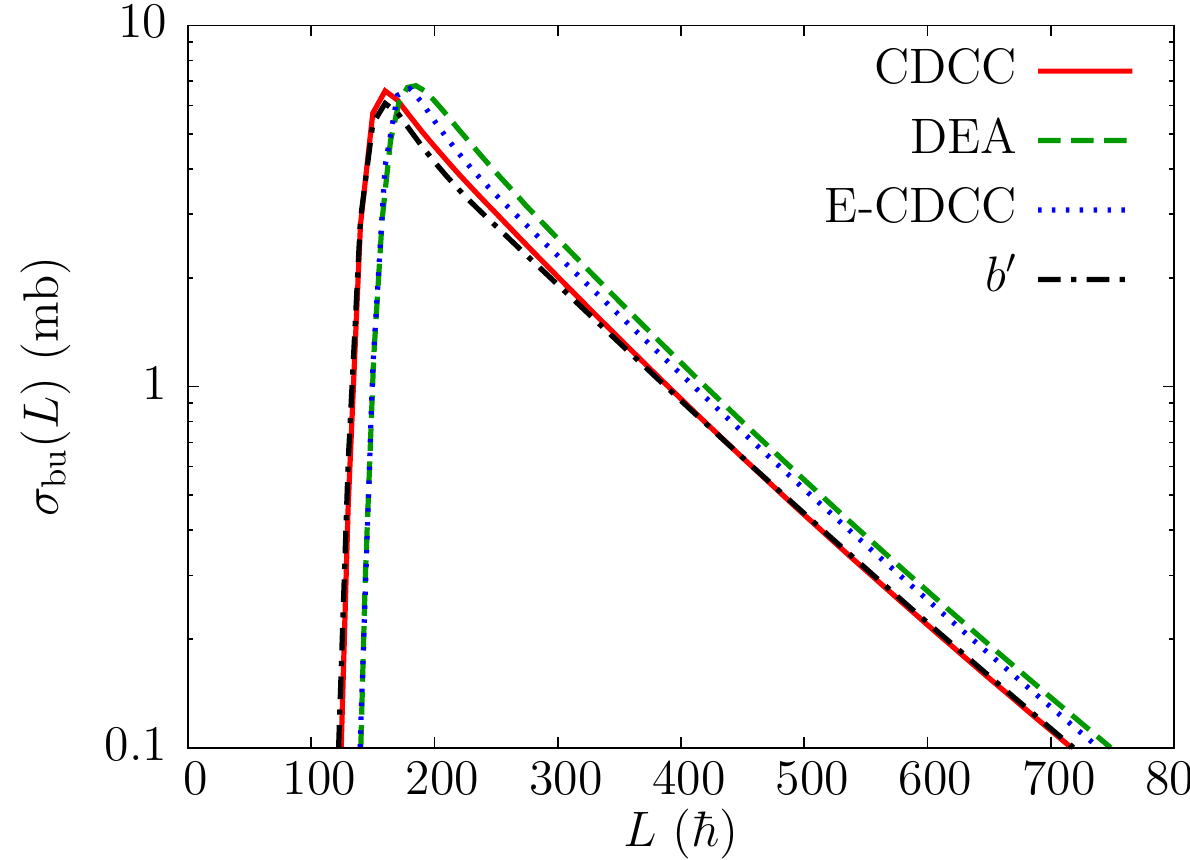}
\caption{Comparison between CDCC and eikonal-based models (E-CDCC and DEA): angular distribution (left) and $L$ contribution to the total breakup cross section (right).
The Coulomb correction \eq{e14} applied to DEA properly accounts for the Coulomb deflection.
The hybrid version of E-CDCC is very accurate although less computational expensive than CDCC.}
\label{f2}
\end{figure}
E-CCDC calculations are compared to full CDCC and DEA results for \ex{15}C on Pb at $20A$MeV in \fig{f2}.
We observe that, unfortunately, it exhibits the same lack of Coulomb deflection as DEA, being also too forward focused.
To better understand this problem, we have looked at the contribution to the total breakup cross section of each angular momentum $L$ for the $P$-$T$ motion (see \fig{f2} right).
We observe that DEA and E-CDCC exhibit very similar results: they seem both shifted towards large $L$ (which corresponds to large $b$ within the eikonal framework) compared to CDCC.
To correct for this, we consider an empirical shift of the transverse component $b$ of the $P$-$T$ relative coordinate $\ve R$.
The idea is to replace $b$ by \cite{BD04}
\beq
b'=\frac{\eta_0}{K_0}+\sqrt{\frac{\eta_0^2}{K_0^2}+b^2},
\eeqn{e14}
which is the distance of closest approach along a classical Coulomb trajectory with impact parameter $b$.
Within DEA, the corresponding results (dash-dotted lines in \fig{f2}) are in excellent agreement with CDCC.

To complete this study, we also analyse the output of the hybrid solution of E-CDCC.
It corresponds to a full CDCC calculation at low $L$ (here below $L_c=500~\hbar$) and an E-CDCC calculation at larger $L$.
It therefore includes all couplings between partial waves where there are significant, while taking the advantage of the eikonal approximation, where it is truly valid.
The result is displayed in \fig{f2} left as a blue dashed line, which is barely visible as it is exactly superimposed to the full CDCC calculation.
This hybrid approach is more efficient that the shift \eq{e14}, however it is also more computationally expensive as it requires some CDCC calculations.

\section{Conclusion}
Breakup reactions are a useful tool to study exotic nuclear structures far from stability such as halo nuclei.
The analysis of experimental data requires accurate reaction models coupled to a realistic description of the projectile.
The most accurate reaction model, CDCC, is very computationally demanding.
Eikonal-based models are simpler to run.
However, they agree with CDCC only at sufficiently high energy (e.g. $70A$MeV).
At lower energy, the eikonal approximation misses the deflection of the incoming projectile by the Coulomb field of the target  \cite{CEN12}.

In this contribution, we show that a way to deal with that is to use a hybrid version in which full CDCC calculations are performed at low $P$-$T$ angular momentum $L$, while the eikonal approximation is used at larger $L$ \cite{Oga03, Oga06}.
This gives a perfect agreement with the full CDCC, although it is much less computationally challenging.

An even simpler solution is to perform an empirical shift.
Albeit not as efficient as the hybrid solution, this shift is more economical in a computational point of view.
It could therefore provide an efficient way to improve the description of the projectile in reaction modelling at a reasonable cost.
One question that remains to be answered is how far down this correction is reliable.
For example, could it be used reliably to analyse experiments performed at low-energy facilities such as HIE-Isolde, or even below?
This would help improving the description of nuclear reactions at such energies, and so the analysis of nuclear structure far from stability.

\section*{Acknowledgment}
This research was supported in part by the Fonds de la Recherche Fondamentale
Collective (grant number 2.4604.07F),
Grant-in-Aid of the Japan Society for the
Promotion of Science (JSPS), and RCNP Young Foreign Scientist
Promotion Program.
This text presents research results of the Belgian Research Initiative on eXotic nuclei
(BRIX), Program No. P7/12, on interuniversity attraction poles of the
Belgian Federal Science Policy Office.



\end{document}